\documentclass[english,aps,pra,showpacs]{revtex4}
\usepackage{graphicx}
\usepackage{babel}
\usepackage{epsf}
\usepackage{amsmath}
\begin{document}
\title{Field-Induced Disorder Point in Non-Collinear Ising Spin Chains}

\author{A.~Vindigni}
\email{alessandro.vindigni@unifi.it}

\affiliation{Universita$'$ di
Firenze Dip. Chimica Inorganica, 3, via della Lastruccia, 50019
Sesto Fiorentino, Italy}

\author{N.~Regnault}
\author{Th.~Jolicoeur}
\email{Nicolas.Regnault, Thierry.Jolicoeur@lpa.ens.fr}

\affiliation{Laboratoire Pierre Aigrain, ENS, D\'epartement de
Physique, 24, rue Lhomond, 75005 Paris, France}

\begin{abstract}
We perform a theoretical study of a non-collinear
Ising ferrimagnetic spin chain inspired by the compound
 Co(hfac)$_2$NITPhOMe. The basic building block of its
structure contains one Cobalt ion and one organic radical each
with a spin 1/2. The exchange interaction is strongly anisotropic
and the corresponding axes of anisotropy have a period three
helical structure. We introduce and solve a model Hamiltonian for
this spin chain. We show that the present compound is very close
to a so-called disorder point at which there is a massive ground
state degeneracy. We predict the equilibrium magnetization
process and discuss the impact of the degeneracy on the dynamical
properties by using arguments based on the Glauber dynamics.
\end{abstract}
\maketitle
\section{Introduction}
Recent advances in coordination chemistry have led to the
synthesis of low-dimensional magnets with unconventional magnetic
properties~\cite{Gatteschi93,Gatteschi94}. Some of them are spin
chains that exhibit complex magnetization behavior under an
applied external field.
The Ising ferrimagnetic spin chain Co(hfac)$_2$NITPhOMe (CoPhOMe
in the
following)~\cite{Caneschi00,Caneschi01,Caneschi02,CoPhOMe_EPL} is
the first quasi-1D compound in which slow dynamics of the
magnetization has been directly observed. For this system it has been
shown that the very long time scale required for establishment of
equilibrium preclude the experimental determination of the static
magnetization curve at low temperature.

This very special
magnet has two kinds of spins S=1/2~: half of them are Cobalt ions
Co$^{2+}$ and the other half are organic radicals NITPhOMe. They
strictly alternate along the chain and have highly anisotropic
exchange interactions best described by an Ising-like Hamiltonian.
The chain itself describes a spiral in real space so that the
local anisotropy axis have a complex pattern~: see Fig.(1). The
primitive cell contains three Cobalt ions alternating with three
radicals. Each Cobalt is related to the other Cobalts by a
120$^\circ$ rotation around the c-axis and has a local anisotropy
axis which makes an angle $\theta$ of approximately 50$^\circ$
with c. The effective S=1/2 spin of the Cobalt and the S=1/2 spin
of the radical have different $g$-values and thus the chain behave
as a unidimensional ferrimagnet due to non-compensation of the
magnetic moments. Only part of the magnetization curve is known
because the exchange energy scale is very high in terms of
accessible magnetic fields. Experiments show a metamagnetic jump
at low fields, much less than the value expected for full
saturation~\cite{Caneschi02,CoPhOMe_EPL}.

In this paper we investigate the equilibrium, static behavior of a
model Hamiltonian for CoPhOMe that we believe encompasses all the
main physical characteristics of this magnet. We show that this
model Hamiltonian can be tuned through a so-called disorder
point~\cite{Stephenson1,Stephenson2} as a function of the
direction $\theta$ of the local axes and magnitude of an applied
external uniform magnetic field. In the case of CoPhOMe, the angle
$\theta$ is not known precisely but our estimates show that it is
very close to the critical value $\theta_c \approx 55^\circ$
expected for the disorder point
($\cos \theta_c =1/\sqrt{3}$, the magic angle of NMR).

In a 1D Ising system with nearest and next-to-nearest neighbor
exchange it is known that under certain
conditions~\cite{Stephenson2}, the spin correlation function can
change from a monotonic exponential decay to a damped oscillating
behavior, passing through zero when the temperature has a special
value called the disorder point. We find that a similar phenomenon
is driven by the geometrical arrangement of anisotropic magnetic
centers even in the sole presence of nearest-neighbor
interactions. The Hamiltonian for CoPhOMe is complicated by the
fact that the eigenvectors of the $g$-tensors do not coincide with
the axis of the chain so that it is not obviously diagonal. We
show that it is possible to choose the quantization axis of the
spins so that all the eigenvalues can be computed exactly. All
thermodynamic properties as well as spin correlations may then be
obtained by a transfer-matrix computation. We are then able to
show that, in a well defined range of  orientation, the
correlation function can switch from a monotonic to an oscillating
behavior by application of a magnetic field. The zero correlation
function condition occurs at a given temperature-dependent field
$h_d(T)$ (disorder field). The monotonic behavior appears when the
ground state is ferrimagnetic while the oscillating behavior is
related to an antiferromagnetic arrangement of the spins.

The disorder point with extensive ground state entropy is
reminiscent of recently observed disorder effects due to non
collinearity in ordered systems (ex. the spin ice residual entropy
in DyTi$_2$O$_7$~\cite{Harris,Sakakibara,Ramirez}).

In  section II we give an analytical solution to the eigenvalue
problem for a ring of $N$ spins; we will show that it is possible
to have a non trivial plateau in the zero temperature
magnetization curve at an arbitrary small field $h_c$. In section
III we study the thermodynamics of the infinite chain obtaining
the geometrical disorder condition in zero field. In section IV we
discuss the consequence of our findings on the dynamical
properties. Finally section V contains our conclusions.

\section{Analytical solution of the eigenvalue problem}

\begin{figure}
\begin{center}
\includegraphics[height=45mm]{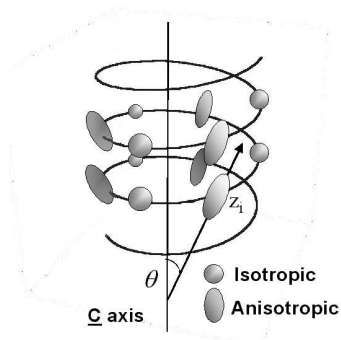}
\caption{The helical structure of the 1D magnet CoPhOMe. Each
anisotropic center has a local axis of anisotropy
$\hat{\mathbf{z}}_i$. These axes are all tilted by a common angle
$\theta$ with respect to the chain axis $\underline{c}$. The angle
between the projections of $\hat{\mathbf{z}}_i$ and
$\hat{\mathbf{z}}_{i+1}$ on the plane perpendicular to
$\underline{c}$ is equal to $\frac{2 \pi}{3}$.}
\label{E:helix}
\end{center}
\end{figure}

\begin{figure}
\begin{center}
\includegraphics[height=45mm]{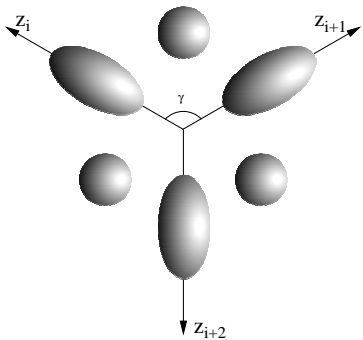}
\caption{The helical structure of CoPhOMe as seen from
the chain axis of symmetry. The anisotropic Cobalt ions
are depicted as ellipsoids and the radical as spheres.}
 \label{Axes}
\end{center}
\end{figure}

In the spin chain  compound CoPhOMe~\cite{Caneschi01,CoPhOMe_EPL}
the $i$th elementary unit cell \textit{for spins} consists of a
pair of
 spins one-half, with Hamiltonian~:
\begin{equation}
\label{E:unit_hamiltonian} \mathcal{H}_i^{\mathrm{intracell}}=
J\,\, {t}^{z_i}_{i}{s}^{z_i}_{i} -\mu_{B}\, g_{iso}\,
\mathbf{B}\cdot {\mathbf{t}}_{i}\, -\mu_{B}\,
 g_{ani}\, (\mathbf{B}\cdot \hat{\mathbf{z}}_i) \, {s}^{z_i}_{i}.
\end{equation}
In this equation, the ${\mathbf{t}}_i$ are spin-1/2 operators
describing the isotropic centers NITPhOMe and the
${\mathbf{s}}_i$ describe the Cobalt S=1/2 anisotropic ions.
For the isotropic organic radicals we have $g_{iso}=2$. The
Ising-like coupling~\cite{Drillon} takes place along the local
axis of anisotropy $\hat{\mathbf{z}}_i$.
The exchange coupling between the radical and the Cobalt is strongly
antiferromagnetic~\cite{Caneschi02}.
The Cobalt ions are
coupled to the external field only by the spin component along the
axis of anisotropy~: the $\hat{\mathbf{z}}_i$ unit vector defines
this direction. The restriction to a ${t}^z_{i}{s}^z_{i}$
Ising coupling, the form of the $g$ tensors and the fact that the
anisotropic centers do not interact directly allow an analytical
treatment of the problem, even when the local directions of
anisotropy $\hat{\mathbf{z}}_i$ are not collinear. Our calculation is
performed for the particular geometry in which the magnetic
centers are arranged in a helical structure with a threefold
periodicity (see  figures \ref{E:helix},\ref{Axes}). In the helical structure
of CoPhOMe, the elementary unit cell of the crystal consists of
three cells of the type described by Equation
(\ref{E:unit_hamiltonian}), with three vectors
$\hat{\mathbf{z}}_1, \hat{\mathbf{z}}_2,\hat{\mathbf{z}}_3$. The
angle of these vectors  with the chain axis $\underline{c}$ is
$\theta$ and the angle between their projections in the plane
perpendicular to $\underline{c}$ is $2\pi/3$ (their vector sum is
along $\underline{c}$).

Since the value of the angle $\theta$ is not known precisely from
experiments, we will solve the problem for arbitrary values of
$\theta$ and obtain constraints in the case of CoPhOMe. From the
study of a monomeric complex of formula Co(hfac)$_2$(NITPhOMe)$_2$
where the Cobalt ion coordination is practically the same as in CoPhOMe,
it is known that $\theta$ is close to 50$^\circ$.

The intracell Hamiltonian  Eq.(\ref{E:unit_hamiltonian}) in zero
field is trivially diagonal on the basis $|m_{iso}\rangle
\otimes|m_{ani}\rangle$ if we take the quantization axis to be the
local axis of the anisotropic center. It is important to note that
this axis is a function of the site and is not fixed over the
whole system. The ${\mathbf{t}}_i$ spin will interact also
with the other nearest neighbor ${\mathbf{s}}_{i+1}$ with the
same Ising Hamiltonian~:
\begin{equation}
\label{E:two_hamiltonian}
\mathcal{H}_{i,i+1}^{\mathrm{intercell}}= J
 {t}^{z_{i+1}}_{i}{s}^{z_{i+1}}_{i+1}.
\end{equation}
For the sake of clarity we have  labelled in
(\ref{E:two_hamiltonian}) the two $z$ axes with the index of their
respective cell. If we want to express all the operators in the
same frame, we can simply choose the crystal frame. Thus
$z_{crys}$ is now the chain axis and a perpendicular
direction $x_{crys}$
can be chosen along the projection of one of the $\hat{\mathbf{z}}_i$ vectors,
say $\hat{\mathbf{z}}_1$.
The simplest way to pass from
the local frame to the crystal frame is to rotate the  tensors~:
\begin{equation}
\underline{J} =\left( \begin{array}{ccc}
0  & 0  & 0 \\
0  & 0  & 0 \\
0  & 0  & J
\end{array}\right),
\,\,\,\,\,\,\,\,\,\,\,\,\,\,\,\,\,\,\,
\underline{g} =\left( \begin{array}{ccc}
0  & 0  & 0 \\
0  & 0  & 0 \\
0  & 0  & g_{ani}
\end{array}\right),
\end{equation}
as ordinary 3x3  matrices. If $\mathcal{R}(\varphi,\theta, \psi)$
is the rotation matrix parameterized by Euler angles that
transforms the cartesian components of a general vector from the
crystal frame to the frame of the $i$th anisotropic center, the
relative $\underline{J}^{i}$ ($\underline{g}^{i}$) tensor is
transformed according to the relation
$\underline{J}\rightarrow\mathcal{R}(\varphi,\theta,\psi =0)$
$\underline{J}$ $\mathcal{R}^{T}(\varphi,\theta,\psi =0)$ (a diagonal
matrix in which the first two terms are equal is insensitive to
the $\psi$ rotation). More explicitly, the transformation
corresponds to a rotation of $\varphi$ about the axis of the helix
and a rotation of $\theta$ about the new y axis. In our particular
geometry,  three successive local $z$ axes are associated to
the values $\varphi=0$, $\frac{2}{3}\pi$, $\frac{4}{3}\pi$ (see
figures \ref{E:helix},\ref{Axes}), in such a way that the effective magnetic
cell of the chain contains six spins. The Hamiltonian expressed in
the crystal frame is then~:
\begin{equation} \label{E:crystal_hamiltonian}
\mathcal{H}=\sum ^{N/2}_{i=1}  {\mathbf{t}}_{i-1}
\cdot\underline{J}^{i}\cdot{\mathbf{s}}_{i} +
{\mathbf{t}}_{i}
\cdot\underline{J}^{i}\cdot{\mathbf{s}}_{i} \,- \mu_{B}  \sum
^{N/2}_{i=1}\,g_{iso}\mathbf{B}\cdot{\mathbf{t}}_{i} +
\mathbf{B}\cdot \underline{g}^{i}\cdot{\mathbf{s}}_{i}
\end{equation}
where the rotated tensors $\underline{J}^{i}$ and
$\underline{g}^{i}$ are periodic with period three with respect to
the site index $i$. We take periodic boundary conditions~: site
${\frac{N}{2}+1}$ is identified to site 1. The Land\'e factor of
the isotropic centers is insensitive to the rotation.

\subsection{Local basis rotation}
To diagonalize the generalized Ising-like Hamiltonian, we first
use the local axis $\hat{\mathbf{z}}_i$ as quantization axis for the Cobalt spin
${\mathbf{s}}_i$. Hence we can replace the spin operator
${\mathbf{s}}_i^z$ by its eigenvalue $\sigma_i/2$,
$\sigma_i =\pm 1$ and we set $\hbar =1$ everywhere.
This is not enough to convert the Hamiltonian to a
diagonal form because the other spins ${\mathbf{t}}_i$ appear
through their projections on \textit{two} special directions~:
$\hat{\mathbf{z}}_i$ as well as $\hat{\mathbf{z}}_{i+1}$ in Eq.(\ref{E:two_hamiltonian}).
However, once we have used this basis we have to solve a set of
decoupled two-dimensional problems since the Hamiltonian is now
given  by~:
\begin{equation}\label{DiagHam}
\mathcal{H}(\{\sigma_i\})=\sum_i \frac{J}{2}[\sigma_i \,
{t}_i^{z_i}+\sigma_{i+1}\, {t}_i^{z_{i+1}}] -\mu_{B}\,
g_{iso}\, \mathbf{B}\cdot{\mathbf{t}}_{i}
-\frac{1}{2}\mu_{B}\, g_{ani}\, \sigma_i\, (\mathbf{B}\cdot
\hat{\mathbf{z}}_i)
\end{equation}
It is an explicit function of the now classical spin configuration
$\{\sigma_i\}$.

We see now that there is an effective magnetic field acting upon
the spins of the radicals but no direct interactions between them~:
\begin{equation}\label{Heff}
\mathcal{H}(\{\sigma_i\}) =\sum_i \mathbf{t}_{i}\cdot
\mathbf{h}_{i}^{\mathrm{eff}} -\frac{1}{2}\mu_{B}\, g_{ani}\,
\sigma_i\, (\mathbf{B}\cdot \hat{\mathbf{z}}_i),
\end{equation}
where the effective field is a function of the applied magnetic
field as well as of the nearest Cobalt spins configuration~:
\begin{equation}\label{Beff}
\mathbf{h}_{i}^{\mathrm{eff}}(\sigma_i,\sigma_{i+1},\mathbf{B}) =
\frac{J}{2}(\sigma_i \, {\hat{\mathbf{z}}_i}+\sigma_{i+1}\,
{\hat{\mathbf{z}}_{i+1}})-\mu_{B}\, g_{iso}\, \mathbf{B}.
\end{equation}
This problem can then be solved immediately~: the eigenenergies of
each radical spin are $\tau_i \mathrm{h}^{\mathrm{eff}}_{i}/2 $
where $\tau_i =\pm 1$ and $\mathrm{h}^{\mathrm{eff}}_{i}$ is the
modulus of the effective field in Eq.(\ref{Beff}). All
eigenenergies of the spin problem are thus given by~:
\begin{equation}\label{Eigen}
E(\{\sigma_i\},\{\tau_i\}) =\frac{1}{2}\sum_i \{\tau_i
\,\,\mathrm{h}^{\mathrm{eff}}_{i}(\sigma_i,\sigma_{i+1},\mathbf{B})
-\mu_{B}\, g_{ani}\, \sigma_i\, (\mathbf{B}\cdot
\hat{\mathbf{z}}_i)\}.
\end{equation}
One has now to deal with a classical spin problem. It is
straightforward to obtain the equilibrium properties of this model
for an arbitrary orientation of the magnetic field. We will
concentrate on the case where the applied field is along the chain
axis since it reveals all the physical features of this problem.
From Eq.(\ref{Eigen}) we see that the ground state spin
configuration is obtained when all variables $\tau_i$ are equal to
-1 for any applied field. The effective field takes only three
distinct values~:
\begin{eqnarray}
\mathbf{h}_{i}^{\mathrm{eff}}(\sigma_i=+1,\sigma_{i+1}=+1,\mathbf{B})
&\equiv& \lambda_{++}=
J\sqrt{\frac{1}{2}(1+\cos\gamma)+g_{iso}^2 (\frac{h}{J})^2 +2g_{iso}\frac{h}{J}\cos\theta}, \\
\mathbf{h}_{i}^{\mathrm{eff}}(\sigma_i=-1,\sigma_{i+1}=-1,\mathbf{B})
&\equiv& \lambda_{--} =
J\sqrt{\frac{1}{2}(1+\cos\gamma)+g_{iso}^2 (\frac{h}{J})^2 -2g_{iso}\frac{h}{J}\cos\theta},\\
\mathbf{h}_{i}^{\mathrm{eff}}(\sigma_i=\pm 1,\sigma_{i+1}=\mp
1,\mathbf{B}) &\equiv& \lambda_{+-} =
J\sqrt{\frac{1}{2}(1-\cos\gamma)+g_{iso}^2 (\frac{h}{J})^2}.\label{BeffVals}
\end{eqnarray}
In these equations, we have defined $h=|\mu_{B}|B$, $\theta$ is the angle between the z-axis and
the vectors $\hat{\mathbf{z}}_i$ and $\gamma$ is the angle between two
consecutive $\hat{\mathbf{z}}_i$ vectors. In the umbrella structure of
CoPhOMe, one has
$\cos\gamma=\frac{3}{2}\cos^{2}\theta-\frac{1}{2}$. To get the
ground state we can equivalently consider the simplified energy~:
\begin{equation}\label{SimplyE}
\tilde{E}(\{\sigma_i\}) =\frac{1}{2}\sum_i \{
-\mathrm{h}^{\mathrm{eff}}_{i}(\sigma_i,\sigma_{i+1},\mathbf{B})
-\mu_{B}\, g_{ani}\, \sigma_i\, (\mathbf{B}\cdot
\hat{\mathbf{z}}_i)\}.
\end{equation}
With only three values for the effective field it can be written
as~:
\begin{equation}\label{SpinEff}
\tilde{E}(\{\sigma_i\}) =-J_{\mathrm{eff}}\sum_i
\sigma_i\sigma_{i+1} -B_{\mathrm{eff}}\sum_i \sigma_i + \mathrm{Cst},
\end{equation}
where~:
\begin{equation}\label{Jeff}
J_{\mathrm{eff}}=\frac{1}{4}\{\lambda_{++}+\lambda_{--}-2\lambda_{+-}
\}\quad \mathrm{and}\quad B_{\mathrm{eff}}=\frac{1}{2}\mu_{B}\,
g_{ani}\, \mathrm{B}\cos\theta
+\frac{1}{4}\{\lambda_{++}-\lambda_{--}\}.
\end{equation}

From these equation it is now clear that the Cobalt spins will
adopt a ferromagnetic (F) or antiferromagnetic (AF) configuration
in the ground state according to the value of the parameter
$J_{\mathrm{eff}}$. If the ground state is AF in zero field,
increasing the applied field will ultimately turn the system to a
F state. Comparing energies, we find that this happens for the
following condition~:
\begin{equation}\label{Cross}
\lambda_{+-}-\lambda_{--}=|\mu_{B}|\, g_{ani}\,
\mathrm{B}\cos\theta .
\end{equation}
(We use conventions for which $\mu_{B} <0$ and the F state at
large field has all $\sigma_i =-1 $).

\begin{figure}
\begin{center}
\includegraphics[height=65mm]  {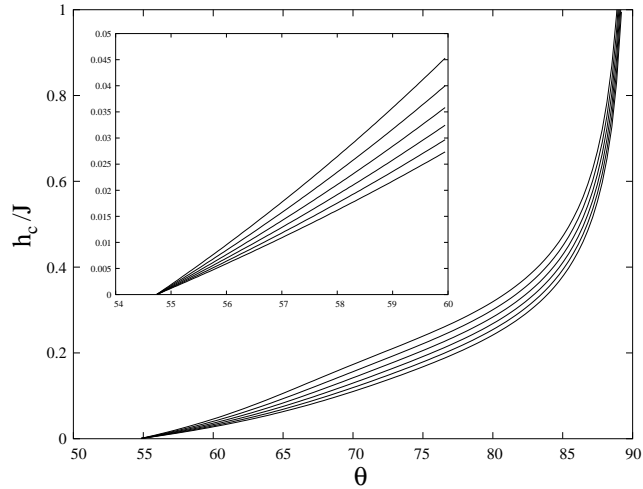}
\caption{The critical field between F and AF ground states as a function
of the angle $\theta$ for $g_{iso}=2$ and $g_{ani}=9.5,9,8.5,8,7.5,7$
from bottom to top.}
\label{Hcrit}
\end{center}
\end{figure}

In zero field, the ground state is ferromagnetic for
$\theta <\theta_{c}$ where $\theta_{c}$ the value
of $\theta$ for which $\cos\gamma=0$~: $\cos\theta_{c}=1/\sqrt{3}$
hence $\theta_{c}\approx 55^\circ$. It is AF for larger values of $\theta$.
Geometrically at the critical angle $\theta_{c}$, the units vectors
$\hat{\mathbf{z}}_i$ are all perpendicular and $\cos\gamma =0$~:
this is the configuration called the "magic angle" in NMR.
If we have a zero-field AF ground state and apply a field
then at some point it will be replaced by the F state~: this will
happen with a jump of the magnetization since the levels do simply cross.
The location of the critical field as a function of the applied field and
the angle $\theta$ is displayed in Fig.(\ref{Hcrit})

For $\theta=\theta_{c}$ we  have a $2^{\frac{N}{2}}$ degenerate
ground state
for all configurations of the Cobalt spins.

\subsection{The $T=0$ Magnetization Curve}
\begin{figure}
\begin{center}
\includegraphics[height=65mm]  {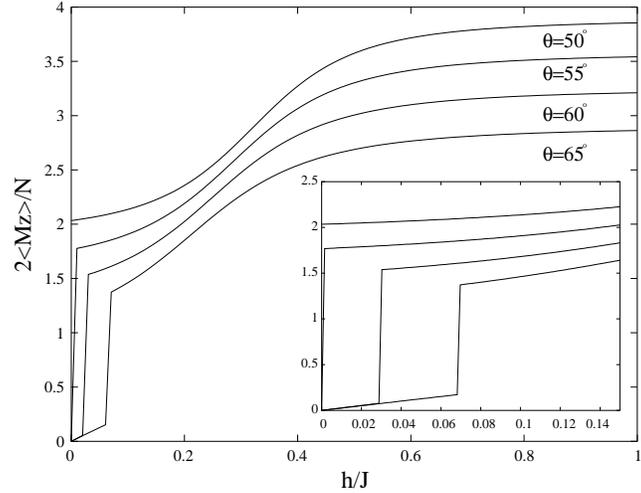}
\caption{Ground state magnetization curve for $\theta=  50^{\circ},\, 55^{\circ},
\,60^{\circ},\, 65^{\circ} $. We use $g_{ani}=9$ and $g_{iso}=2$
as appropriate for CoPhOMe.}
\label{GSMag}
\end{center}
\end{figure}
We now discuss the magnetization curve at zero temperature of the previous model.
From the ingredients of the previous section, it is straightforward
to derive simple expressions for the magnetization. In the ground state
we have the sum of the contribution of the Cobalt spins that are always completely
polarized along their local anisotropy axis and the contribution from the
radical spins that always follow exactly the local field~:
\begin{equation}\label{ExactMag}
\langle \mathbf{M} \rangle =\sum_i\frac{1}{2}\,
\mu_{B}\, g_{ani}\,\sigma_i\,\hat{\mathbf{z}}_i
-\frac{1}{2}\,\mu_{B}\, g_{iso}\,
\frac{\mathbf{h}_{i}^{\mathrm{eff}}}{\|\mathbf{h}_{i}^{\mathrm{eff}}\|}.
\end{equation}
This expression is valid for an arbitrary spin configuration. In the F state
the magnetization is nonzero only along the z axis and equal to~:
\begin{equation}\label{MF}
\frac{2}{N}\langle M^{z} \rangle =-\frac{1}{2}\,
\mu_{B}\, g_{ani}\,\cos\theta
+\frac{1}{2}\,\mu_{B}\, g_{iso}\,
\frac{J\cos\theta +\mu_{B}\, g_{iso}\, B}{\lambda_{--}},
\end{equation}
N is the total number of spins Cobalt + radical.
In the AF state, the magnetization from Eq.(\ref{ExactMag}) is zero
in zero applied field and again only along z in nonzero field.
We find then~:
\begin{equation}\label{MAF}
\frac{2}{N}\langle M^{z} \rangle =\frac{1}{2}\,
\mu_{B}^2\, g_{iso}^2\,
\frac{ B}{\lambda_{+-}}.
\end{equation}
Sample curves are displayed in Fig.(\ref{GSMag}). For F ground states
the magnetization process is smooth and converges to a saturation value
$ (g_{ani}\,\cos\theta +g_{iso})|\mu_{B}|/2$ which is reached however
only at $B\rightarrow\infty$. In the AF case, for $\theta$ large enough
there is a jump of the magnetization when the spins reorients
themselves into the ferromagnetic state.

From the shape of the
known magnetization curve of CoPhOMe, we find that we need an angle
$\theta$ which is close to the critical value.
From the saturation value we get an estimate of
the $g$-factor of the Cobalt ions once we assume that the $g$ of the radical
is extremely close to 2. Considering present experimental data~\cite{CoPhOMe_EPL},
the preferred values are $g_{ani}\simeq 9$ and $\theta \simeq 54^\circ$.


\section{Finite Temperature properties}
The formula (\ref{Eigen}) we have found for the
eigenvalues of the closed ring can be used to obtain the finite
temperature properties, in the thermodynamic limit, with the
transfer matrix method. The analytical treatment will be given
only for the field applied along the chain axis; in this case two
successive anisotropic spins experience the same field and the
magnetic unit cell  consists of one isotropic-anisotropic pair.

\subsection{The $\sigma$-$\sigma$ Transfer Matrix}
If we explicitly sum over the $\underline{\tau}$  coordinates we can
define the $\sigma$-$\sigma$ transfer matrix~:
\begin{equation} \label{E:Transfer_Matrix}
T_{\sigma, \sigma'}=\exp \Bigl[- \frac{1}{4}\beta\,
g_{ani}\, h \, \cos\theta\,(\sigma+\sigma') \Bigr]
\sum_{\tau =\pm1}  \exp \Bigl[ \frac{\beta}{2}
\tau\,\mathrm{h}^{\mathrm{eff}}(\sigma,\sigma',B) \Bigr],
\end{equation}
and obtain the well known expression for the partition function of
the problem~:
\begin{equation}
\mathcal{Z}=\mathcal{T}r \left[ T_{\sigma, \sigma'} ^{\frac{N}{2}} \right] =
\Lambda_{+}^{\frac{N}{2}} + \Lambda_{-}^{\frac{N}{2}}
\end{equation}
where  $\Lambda_{+}$ and  $\Lambda_{-}$ are the eigenvalues of the matrix
$T_{\sigma, \sigma'}$. In the thermodynamic limit the largest
eigenvalue dominates and $\mathcal{Z}=\Lambda_{+}^{\frac{N}{2}}$.
The spin correlation function for $N\rightarrow \infty$ is given
by~:
\begin{equation} \label{E:pairspincorr}
\langle\sigma_{k}\sigma_{l}\rangle = |\langle\phi^{+}|\sigma|\phi^{+}\rangle|^{2} +
 |\langle\phi^{+}|\sigma|\phi^{-}\rangle|^{2}
 \Bigl(\frac{\Lambda_{-}}{\Lambda_{+}}\Bigr)^{|k-l|},
\end{equation}
 where the $|\phi^{\pm}\rangle$ are the two eigenvectors of the transfer matrix (\ref{E:Transfer_Matrix}).
\subsection{The Zero-Field Spin Correlation Function}
The first term in (\ref{E:pairspincorr}) is simply the square
expectation value of each anisotropic spin  and it is different
from zero only in non zero field. In zero field the solution to
the eigenvalues problem  gives~:
\begin{equation}
\Lambda_{\pm} = 2[\cosh(\nu_{\uparrow\uparrow})  \pm
\cosh(\nu_{\uparrow\downarrow})]  \quad \mathrm{and}\quad
|\phi^{\pm}\rangle=\frac{1}{\sqrt{2}} \binom{1}{\pm1},
\end{equation}
where~:
\begin{equation}\label{Defs}
\nu_{\uparrow\uparrow} = \frac{1}{2}\beta\lambda_{++},
\,\,\,\,\,\,\,\,\,\,\,\,\,\, \nu_{\uparrow\downarrow} =
\frac{1}{2}\beta\lambda_{+-},
\,\,\,\,\,\,\,\,\,\,\,\,\,\, \nu_{\downarrow\downarrow}=
\frac{1}{2}\beta\lambda_{--},
\end{equation}
and  in zero field $\nu_{\uparrow\uparrow}
=\nu_{\downarrow\downarrow}$. The $\sigma$-$\sigma$  correlation
function is~:
\begin{equation} \label{E:pairspincorr_zf}
\langle\sigma_{k}\sigma_{l}\rangle =
\biggl[\frac{\cosh(\nu_{\uparrow\uparrow}) -
\cosh(\nu_{\uparrow\downarrow})} {\cosh(\nu_{\uparrow\uparrow}) +
\cosh(\nu_{\uparrow\downarrow})} \biggr]^{|k-l|},
\end{equation}
that decays exponentially with increasing spin separation for
every non zero temperature.

For  $\nu_{\uparrow\uparrow} > \nu_{\uparrow\downarrow}$
($\theta<\theta_{c}$) the decay of (\ref{E:pairspincorr_zf}) is
monotonic whereas for   $\nu_{\uparrow\uparrow} <
\nu_{\uparrow\downarrow}$ ($\theta>\theta_{c}$) it is oscillatory
with sign $(-1)^{|k-l|} $. For $\theta=\theta_{c}$  the expression
(\ref{E:pairspincorr_zf}) vanishes at any  temperature; this is
related to the fact that the energies  (\ref{Eigen}) do not depend
on the $\underline{\sigma}$  coordinates. The thermal fluctuations
lead to the same behavior in the Ising ferromagnet with  an
antiferromagnetic next-to-nearest neighbor interaction, but not
strong enough to destroy the ferromagnetic structure of the ground
state~\cite{Stephenson1,Stephenson2}. Here it is the $\theta$
angle which is the tuning parameter instead of the temperature
that makes the system pass through a \textit{disorder point} at
$\theta=\theta_{c}$. From the Eq.(\ref{E:pairspincorr_zf}) we can
also  obtain an expression for the zero-field correlation length~:
\begin{equation} \label{E:corr_length_zf}
\xi^{-1}=-\frac{1}{2}\ln\biggl|\frac{\cosh(\nu_{\uparrow\uparrow})
- \cosh(\nu_{\uparrow\downarrow})} {\cosh(\nu_{\uparrow\uparrow})
+ \cosh(\nu_{\uparrow\downarrow})} \biggr|.
\end{equation}
The factor one half comes from the fact that the correlation function
Eq.(\ref{E:pairspincorr_zf})
corresponds to an actual separation of $2|k-l|$  magnetic centers.

\begin{figure}
\begin{center}
  \leavevmode
  \vspace{0.1cm}
  \hspace{-0.5cm}
  \epsfxsize=8cm \epsfbox{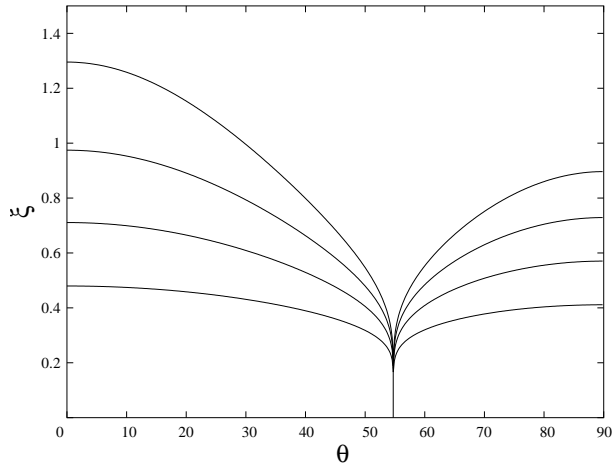}
\caption{Dependence upon the angle $\theta$ of the Cobalt
correlation length $\xi$ in zero field for  temperatures
given by $J/T = 2,1.5,1,0.5$ from top to bottom.}
\label{xi_zero_field}
\end{center}
\end{figure}

The Eq.(\ref{E:corr_length_zf}) gives back the Ising model
correlation length when $\theta=0$. From  figure
(\ref{xi_zero_field})  we observe that at any temperature while
increasing  $\theta$  the correlation length decreases until it
reaches $\theta_{c}$. When $\theta > \theta_{c}$  it increases
again, but in this case the effective coupling between the
${\mathbf{s}}_i$ spins is antiferromagnetic. For
$\theta=\theta_{c}$ it vanishes, as well as the correlation
function, at all temperatures.

\subsection{Field dependence of the critical point}

It is straightforward to wonder if it is  possible to have a
vanishing correlation length for $\theta\ne\theta_{c}$ using the
external field as a tunable parameter. This happens when the
smallest eigenvalue of the transfer matrix is zero (see equation
(\ref{E:pairspincorr})). The eigenvalues in presence of the field are~:
\begin{equation} 
\Lambda_{\pm} = X\cosh(\nu_{\uparrow\uparrow}) + X^{-1}
\cosh(\nu_{\downarrow\downarrow})\pm \sqrt{ \bigl[
X\cosh(\nu_{\uparrow\uparrow}) -
X^{-1}
\cosh(\nu_{\downarrow\downarrow}) \bigr]^{2}
 +4  \cosh^{2}(\nu_{\uparrow\downarrow}) },
\end{equation}
where~:
\begin{equation}\label{Auxiliary}
    X=\exp (-\frac{\beta}{2}g_{ani}h\cos\theta )
\end{equation}
The $\Lambda_{-}=0$ condition is fulfilled when~:
\begin{equation} \label{E:disorder_field_condition}
\cosh(\nu_{\uparrow\uparrow}) \cosh(\nu_{\downarrow\downarrow}) =
\cosh^{2}(\nu_{\uparrow\downarrow}).
\end{equation}
This equation can be solved numerically for any  $\theta$ in order to obtain what we
 call the \textit{disorder field} $h=h_{d}$
for which~:
\begin{equation}
\langle\sigma_{k}\sigma_{l}\rangle-\langle\sigma_{k}\rangle^{2}=
\exp\biggl[-\frac{2|k-l|}{\xi}\biggr]  = 0.
\end{equation}
A numerical analysis shows that the Eq.(\ref{E:disorder_field_condition})
has a solution only for $0
< \theta \leq \theta_{c}$. In the Ising limit ($\theta=0$) the
equation (\ref{E:disorder_field_condition}) reduces to~:
\begin{equation} 
\cosh^{2}(2\beta J) - \tanh^{2}(\beta g_{iso} h/2) \sinh^{2}(2\beta
J)=1,
\end{equation}
that has no solution for any  $h<\infty$. On the other hand we
know that for $\theta=\theta_{c}$ the  disorder field is
$h_{d}=0$. Moreover the Eq.(\ref{E:disorder_field_condition}) is
symmetric in the exchange $h \leftrightarrow-h$. Since there is no
physical reason to think the $h_{d}$ dependence on $\theta$ to be
discontinuous this means that the  disorder field already takes all
the allowed values in the range $0 < \theta \leq \theta_{c}$,
leaving no solution to the Eq.(\ref{E:disorder_field_condition}) for
$\theta > \theta_{c}$. This is in agreement with  the numerical
results.

\begin{figure}
\begin{center}
  \leavevmode
  \vspace{0.1cm}
  \hspace{-0.5cm}
  \epsfxsize=10cm \epsfbox{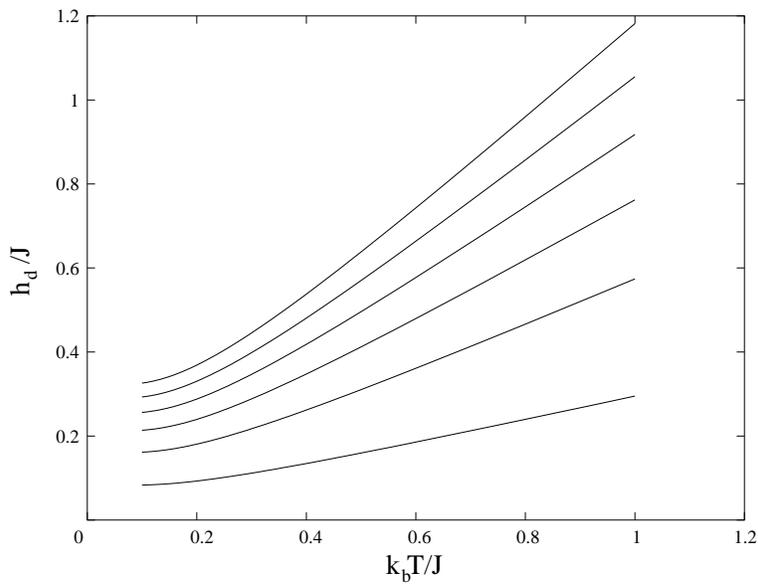}
  \caption{Disorder  field dependence on  $T$ for $\theta= 44^{\circ},
 46^{\circ}, 48^{\circ}, 50^{\circ}, 52^{\circ}, 54^{\circ} $ from top to bottom.}
\label{E:criticalfield_theta_T}
\end{center}
\end{figure}

In  figure (\ref{E:criticalfield_theta_T}) the disorder  field
dependence on the temperature is shown for different values of
$\theta$. This dependence is less strong for $ \theta$ approaching
 $ \theta_{c}$. For  $\theta=\theta_{c}$, we
find a line lying on the zero-field  axis.

\begin{figure}
\begin{center}
  \leavevmode
  \vspace{0.1cm}
  \hspace{-0.5cm}
  \epsfxsize=10cm \epsfbox{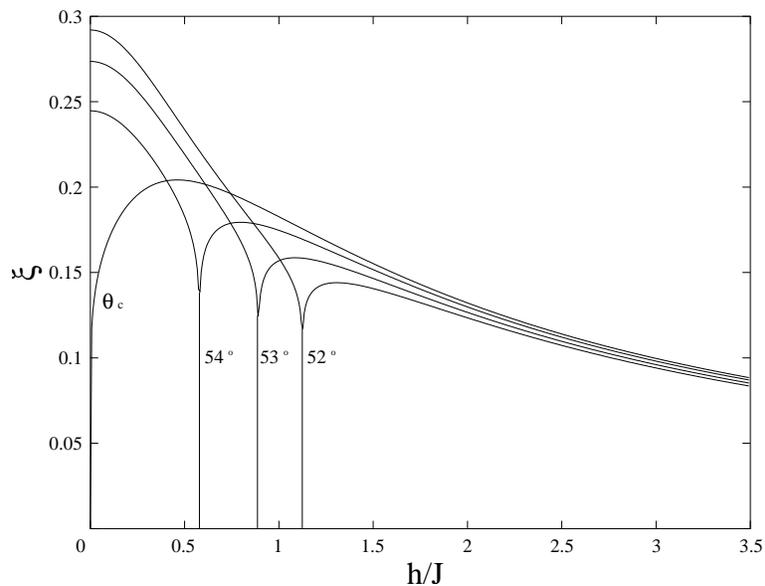}
  \caption{\small Correlation length as a function of  $h/J=|\mu_{B}|B/J$
  for $\frac{J}{T}=0.5$ and
$\theta=52^{\circ}, 53^{\circ},  54^{\circ}, \theta_{c}$.}
\label{E:xi_field}
\end{center}
\end{figure}

In figure (\ref{E:xi_field}) we note that the critical point
appears as cusp in the correlation length $vs$ field plot. If we
consider first  the case of $\theta < \theta_{c}$, when the field
is switched on,  the correlation length decreases because the
fluctuations of the spins are progressively damped by the effect
of the field; in this region the nearest Cobalt spins  are
ferromagnetically correlated ($
\langle\sigma_{k}\sigma_{k+1}\rangle-\langle\sigma_{k}\rangle^{2}>0
$). Then $\xi$ abruptly drops to zero for $h=h_{d}$, at the
critical point. For greater fields the fluctuations are
antiferromagnetically correlated ($
\langle\sigma_{k}\sigma_{k+1}\rangle-\langle\sigma_{k}\rangle^{2}<0
$), even if the field orders the Cobalt ions in a relative
ferromagnetic arrangement. It is because of this increasing order
that the correlation length, a part from the narrow region of the
cusp, smoothly goes to zero for $h\rightarrow \infty$.

In the case of $\theta = \theta_{c}$ the cusp lies at the origin
and so the $\sigma$-$\sigma$ fluctuations in presence of the field
are always antiferromagnetically correlated. From this point of
view, in this particular geometry,  the behavior is similar to the
one we have for  $\theta > \theta_{c}$. It is worth noticing that
for $\theta > \theta_{c}$ the sign of the fluctuations is always
coherent with the structure of the zero-field ground  state, while
for $\theta = \theta_{c}$ the zero-field  expectation value of
each ${\mathbf{s}}_{i}$  spin vanishes.

\section{Stochastic dynamics}

In the last section we have seen that many different behaviors of
the correlation length appear by varying  the geometrical
($\theta$) and the external parameters ($T$ and $h$) of the
system. The most peculiar feature of the real compound CoPhOMe is
the slow relaxation of the magnetization on a macroscopic scale
\cite{Caneschi01}. This behavior has been explained up to now only
through a modified Glauber model \cite{CoPhOMe_EPL} that was based
only on a simplified Hamiltonian that did not contain all the
complexity of Eq.(\ref{E:crystal_hamiltonian}). From a theoretical
point of view the correlation length and the slowest time scale
are usually related by mean of a dynamical critical exponent. Thus
we can easily imagine this system to show a very rich scenario in
the  long time scale dynamics too. In this section we investigate
the implications of detailed balance on the transition functions
$W(\sigma_{k})$  and $W(\tau_{k})$ involved in a Glauber-like
approach~\cite{Glauber}. The Eq.(\ref{Eigen}) in zero field can be
written as~:
\begin{equation} \label{E:ring_eigenvalues_dyn}
\beta \mathcal{E(\underline{\sigma}, \underline{\tau})}=\sum
^{N/2}_{i=1}   \tau_{i}\, \biggl[
\frac{\nu_{\uparrow\uparrow}+\nu_{\uparrow\downarrow}}{2} +
\sigma_{i}\sigma_{i+1}
\frac{\nu_{\uparrow\uparrow}-\nu_{\uparrow\downarrow}}{2} \biggr].
\end{equation}
The ratio of the two equilibrium probabilities corresponding to
the configurations $(\sigma_{1}, \tau_{1},\ldots  \sigma_{k},
\tau_{k},\sigma_{k+1}, \ldots  \sigma_{\frac{N}{2}},
\tau_{\frac{N}{2}})$ and $(\sigma_{1}, \tau_{1},\ldots \sigma_{k},
-\tau_{k},\sigma_{k+1}, \ldots  \sigma_{\frac{N}{2}},
\tau_{\frac{N}{2}})$ is~:
\begin{equation} \label{E:equilibrium_radical_ratio}
\frac{P_{eq}(-\tau_{k}) }{ P_{eq}(\tau_{k}) } = \frac{
\exp\Bigl(\tau_{k}\frac{\nu_{\uparrow\uparrow}+\nu_{\uparrow\downarrow}}{2}\Bigr)
} {
\exp\Bigl(-\tau_{k}\frac{\nu_{\uparrow\uparrow}+\nu_{\uparrow\downarrow}}{2}\Bigr)
} \,\, \frac{ 1 + \tau_{k}\sigma_{k}\sigma_{k+1}
\tanh\Bigl(\frac{\nu_{\uparrow\uparrow}-\nu_{\uparrow\downarrow}}{2}\Bigr)
} {   1 - \tau_{k}\sigma_{k}\sigma_{k+1}
\tanh\Bigl(\frac{\nu_{\uparrow\uparrow}-\nu_{\uparrow\downarrow}}{2}\Bigr)
}.
\end{equation}
which $must$ be equal, for the detailed balance condition to be
satisfied, to the ratio $\frac{W(\tau_{k})}{W(-\tau_{k})}$. The
(\ref{E:equilibrium_radical_ratio}) tells us that even if the
energy gap between the magnetic and the non magnetic  state can be
arbitrarily small for $\theta\sim \theta_{c}$
($\nu_{\uparrow\uparrow} \sim \nu_{\uparrow\downarrow}  $),  the
energy barrier appearing in the transition probability
$W(\tau_{k})$ will always  be $\Delta E \simeq
\frac{1}{2}(\nu_{\uparrow\uparrow} + \nu_{\uparrow\downarrow})$,
that does not vanish for any $\theta$.

For the sake of completeness, we write the equivalent of the
(\ref{E:equilibrium_radical_ratio}) for the Cobalt spins~:
\begin{equation}
\frac{P_{eq}(-\sigma_{k}) }{ P_{eq}(\sigma_{k}) } = \frac{1 +
\frac{1}{2} \sigma_{k} (\tau_{k}\sigma_{k+1} +
\tau_{k-1}\sigma_{k-1})
\tanh(\nu_{\uparrow\uparrow}-\nu_{\uparrow\downarrow})} { 1 -
\frac{1}{2} \sigma_{k} (\tau_{k}\sigma_{k+1} +
\tau_{k-1}\sigma_{k-1})
\tanh(\nu_{\uparrow\uparrow}-\nu_{\uparrow\downarrow})}.
\end{equation}
In this case we see that the ratio $\frac{P_{eq}(-\sigma_{k}) }{
P_{eq}(\sigma_{k}) }\rightarrow 1$ for   $\theta \rightarrow
\theta_{c}$.

Even if  we do not have so far a rigorous treatment for the
dynamics of this system, it seems reasonable to relate the
quantity $\Delta E$ to the activation energy of the relaxation
process. For our purpose it is enough to observe that the presence
of a jump in the $T=0$ magnetization curve, occurring at
relatively small fields, and the presence of a tunable disorder
point is consistent with the observation of the slow decay of the
macroscopic magnetization.

\section{Conclusion}

We have introduced a magnetic Hamiltonian suited to the
description of the Ising-like spin chain CoPhOMe. This model
contains two kinds of spins with very different $g$-factors. We
have shown that by a careful choice of quantization axis all
eigenvalues can be determined analytically if the applied field is
along the axis of the chain. As a consequence we have obtained the
magnetization curve. This curve displays a metamagnetic jump at
some critical value of the external applied field. The position of
the jump is closely related to the presence of a disorder point of
the magnetic model. This disorder point can be reached
theoretically by tuning the angle of the Cobalt anisotropy axis
with the chain axis. All experimental data indicate that the real
chain CoPhOMe is very close to this critical value. Right at this
disorder point there is extensive ground state degeneracy. By
using ideas taken from the Glauber dynamics we show that this is
likely to be related to the very long relaxation times observed
when performing magnetization measurements on this chain. There is
still the problem of developing a realistic model of the dynamical
properties of this spin chain.

\end{document}